\documentclass[letterpaper, 11 pt, journal, onecolumn]{IEEEtran}
\usepackage{setspace}
\pdfoutput=1
\usepackage{hyperref}
\hypersetup{hidelinks,pdfnewwindow,breaklinks} 
\usepackage[ruled,vlined]{algorithm2e}
\usepackage{color}
\usepackage{xcolor}
\usepackage{amsmath}
\usepackage{nomencl}
\usepackage{amssymb}
\usepackage{amsfonts}
\usepackage{graphicx}
\usepackage{bm}
\usepackage{bbm}
\usepackage{import}
\usepackage{mathrsfs}
\usepackage{array}
\usepackage{mathtools}
\mathtoolsset{showonlyrefs=true} 
\usepackage{siunitx}
\usepackage{calligra}
\usepackage{makecell}
\usepackage{cuted}
\usepackage{rotating} 
\usepackage{marginnote}
\usepackage{caption}
\usepackage{subcaption}
\usepackage{cite}
\usepackage{scalerel}
\usepackage{boxedminipage}

\makeatletter
\renewcommand{\@begintheorem}[2]{%
  \trivlist
  \item[\hskip\labelsep
  {\indent\color{ieeeblue}\itshape #1\ #2.}]%
  \itshape
}

\renewcommand{\@opargbegintheorem}[3]{%
  \trivlist
  \item[\hskip\labelsep
  {\indent\color{ieeeblue}\itshape #1\ #2\ (#3).}]%
  \itshape
}
\makeatother
\newtheorem{theorem}{Theorem}
\newtheorem{lemma}{Lemma}
\newtheorem{proposition}{Proposition}
\newtheorem{corollary}{Corollary}
\newtheorem{assumption}{Assumption}

\newtheorem{remark}{Remark}
\newtheorem{definition}{Definition}

\newcommand{\Xcal}{\mathcal{X}}
\newcommand{\Ucal}{\mathcal{U}}
\newcommand{\Prob}{\mathbb{P}}

\newcommand{\Xonew}{x_{0,i_{\text{new}}}}
\newcommand{\XXnew}{X_{i_{\mathrm{new}}}}
\newcommand{\YYnew}{Y_{i_{\mathrm{new}}}}

\newcommand{\Strain}{D_{\mathrm{train}}}
\newcommand{\Dcal}{D_{\mathrm{cal}}}
\newcommand{\Dcald}{D_{\mathrm{cal}}^\Delta}
\newcommand{\Dcalr}{D_{\mathrm{cal}}^\rho}
\newcommand{\Ntrain}{N_{\mathrm{train}}}
\newcommand{\Ncal}{N_{\mathrm{cal}}}
\newcommand{\Itrain}{\mathcal{I}_{\mathrm{train}}}
\newcommand{\Ical}{\mathcal{I}_{\mathrm{cal}}}
\newcommand{\Icald}{\mathcal{I}_{\mathrm{cal}}^\Delta}
\newcommand{\Icalr}{\mathcal{I}_{\mathrm{cal}}^\rho}
\newcommand{\Scored}{S_i^\Delta}
\newcommand{\Scorer}{S_i^\rho}
\newcommand{\deltad}{\delta^\Delta}
\newcommand{\deltar}{\delta^\rho}
\newcommand{\quantile}{\mathcal{Q}}
\newcommand{\quantiled}{\mathcal{Q}^\Delta}
\newcommand{\quantiler}{\mathcal{Q}^\rho}
\newcommand{\myhref}[2]{\textcolor{uiucbluedark}{\href{#1}{#2}}}

\definecolor{caltechblue}{rgb}{0,0.2314,0.5980}

\definecolor{ieeeblue}{HTML}{004293}
\definecolor{uiucbluedark}{HTML}{004293}
\definecolor{subsectioncolor}{rgb}{0,0.541,0.855}
\definecolor{nblue}{HTML}{004293}

\makeatletter
\def\section{\@startsection{section}{1}{\z@}
{3.0ex plus 1.5ex minus 1.5ex}
{0.7ex plus 1ex minus 0ex}
{\color{uiucbluedark}\centering\normalfont\normalsize\scshape}}

\def\subsection{\@startsection{subsection}{2}{\z@}
{3.5ex plus 1.5ex minus 1.5ex}
{0.7ex plus .5ex minus 0ex}
{\color{subsectioncolor}\normalfont\normalsize\itshape\raggedright}}

\def\subsubsection{\@startsection{subsubsection}{3}{\parindent}
{0ex plus 0.1ex minus 0.1ex}
{0ex}
{\color{uiucbluedark}\normalfont\normalsize\itshape}}
\makeatother

\makeatletter
\def\ps@IEEEtitlepagestyle{
\def\@oddhead{\hbox{}\@IEEEheaderstyle\leftmark\hfil\thepage}\relax
\def\@evenhead{\@IEEEheaderstyle\thepage\hfil\leftmark\hbox{}}\relax
  \def\@oddfoot{\mycopyrightnotice}
  \def\@evenfoot{}
}

\def\mycopyrightnotice{
  {\scriptsize\sffamily
  \begin{boxedminipage}{\textwidth}
  \centering
  © 2025 IEEE. Personal use of this material is permitted. Permission from IEEE must be obtained for all other uses, in any current or future media, including reprinting/republishing this material for advertising or promotional purposes, creating new collective works, for resale or redistribution to servers or lists, or reuse of any copyrighted component of this work in other works (DOI: \myhref{https://doi.org/10.1109/LCSYS.2025.3578062}{10.1109/LCSYS.2025.3578062}).
  \end{boxedminipage}
  }
}

\def\BibTeX{{\rm B\kern-.05em{\sc i\kern-.025em b}\kern-.08em
    T\kern-.1667em\lower.7ex\hbox{E}\kern-.125emX}}
\title{\LARGE \bf~~~\\\vspace{0.1em}
\textcolor{uiucbluedark}{Statistical Guarantees in Data-Driven Nonlinear Control:\\Conformal Robustness for Stability and Safety (Extended Version)}
}
\author{\normalsize{Ting-Wei Hsu and Hiroyasu Tsukamoto}
\thanks{\scriptsize\sffamily Part of the research was carried out at the Jet Propulsion Laboratory, California Institute of Technology, under a contract with the National Aeronautics and Space Administration. Email Addresses: \myhref{mailto:twhsu3@illinois.edu}{twhsu3@illinois.edu} \& \myhref{mailto:hiroyasu@illinois.edu}{hiroyasu@illinois.edu}.}
\\
\textit{Department of Aerospace Engineering, University of Illinois Urbana-Champaign}
}

\begin{document}
\maketitle
\vspace{-2em}

\begin{abstract}
We present a true-dynamics-agnostic, statistically rigorous framework for establishing exponential stability and safety guarantees of closed-loop, data-driven nonlinear control. 
Central to our approach is the novel concept of \emph{conformal robustness}, which robustifies the Lyapunov and zeroing barrier certificates of data-driven dynamical systems against model prediction uncertainties using conformal prediction in a \emph{closed loop}. 
It quantifies these uncertainties by leveraging rank statistics of prediction scores over system trajectories, without assuming any specific underlying structure of the prediction model or distribution of the uncertainties. 
With the quantified uncertainty information, we further construct the conformally robust control Lyapunov function (CR-CLF) and control barrier function (CR-CBF), data-driven counterparts of the CLF and CBF, for fully data-driven control with statistical guarantees of finite-horizon exponential stability and safety.
The performance of the proposed concept is validated in numerical simulations with four benchmark nonlinear control problems. 
\end{abstract}


\section{Introduction}
Data-driven control is reshaping how autonomous systems operate under uncertainties, enabling decision-making without explicit knowledge of true dynamics. Yet, a formal approach to establishing their performance guarantees remains non-standardized, whereas existing methods often rely on \textcolor{uiucbluedark}{\textbf{(i)}}~knowledge of underlying dynamics, \textcolor{uiucbluedark}{\textbf{(ii)}} structural constraints in prediction models,  or \textcolor{uiucbluedark}{\textbf{(iii)}} assumptions on probability distributions or structures of system uncertainties.

\subsubsection*{Contributions}
\begin{figure}[htbp]
    \centering
    \includegraphics[width=0.9\textwidth]{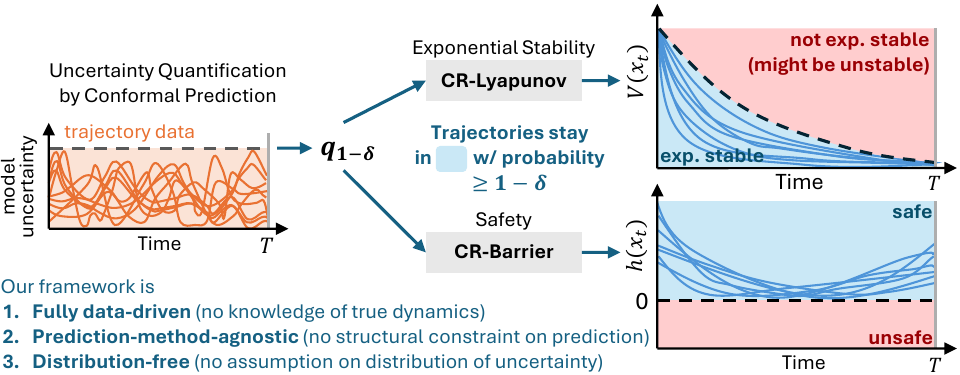}
    \caption{Illustration of our proposed framework.}
    \label{fig:intro}
\end{figure}

We propose \emph{conformal robustness}, a statistically rigorous concept for robustifying finite-horizon stability and safety guarantees in data-driven nonlinear dynamical systems.
Given any data-driven prediction models of the system dynamics, we quantify the model uncertainties over a finite time horizon using conformal prediction \cite{2022_Vovk_CP-book,2008_Shafer_CP-tutorial,2021_Angelopoulos_CP-Intro,2024_Lindemann_CP-control-survey}.
We then construct the conformally robust control Lyapunov function (CR-CLF) and control barrier function (CR-CBF) by robustifying the standard CLF and CBF conditions using the quantified uncertainty information.
This enables systematic \emph{closed-loop} control designs with provable finite-horizon stability and safety guarantees in data-driven nonlinear systems. 
Notably, unlike most existing methods, our proposed framework is fully data-driven and works without the need for \textcolor{uiucbluedark}{\textbf{(i)}}~--~\textcolor{uiucbluedark}{\textbf{(iii)}}.
Our proposed framework is illustrated in Fig.~\ref{fig:intro}.

In simulations with four benchmark nonlinear control problems, we demonstrate the CR-CLFs and CR-CBFs in fully data-driven settings. We also show that they can be synthesized by quadratic programs and neural networks. \newpage

\subsubsection*{Related Work}
Control Lyapunov functions (CLFs) \cite{1983_Sontag_CLF, 1983_Artstein_CLF, 1996_Freeman} and control barrier functions (CBFs) \cite{2007_Wieland_OG-CBF, 2014_Ames_CBF-QP-ACC, 2017_Ames_CBF-QP} are among the most widely studied and applied tools for providing formal stability and safety guarantees in safety-critical nonlinear dynamical systems.
They typically require partial knowledge of the true dynamics, and are not directly applicable to fully data-driven models with uncertainties. 
Most prior works assume specific underlying structures or probability distributions of the uncertainties.
For instance, robust CLFs \cite{1996_Freeman} and CBFs \cite{2018_Jankovic_Robust-CBF} require a known upper bound on the uncertainties.
Stochastic CLFs \cite{1996_Florchinger_SCLF} and CBFs \cite{2021_Clark_SCBF} model the uncertainties using Wiener processes. 
Other approaches may impose specific structural assumptions on the uncertainties in the CLFs \cite{2019_Taylor_CLF-Learn} and CBFs \cite{2020_Taylor_CBF-Learn} and utilize machine learning to learn such uncertainties.
Yet, the development of fully data-driven frameworks for constructing stability and safety guarantees that are robust against model uncertainties---with no structural or distributional assumptions---remains relatively unexplored.

Data-driven control has then attracted significant attention from the statistical learning and control community, as an alternative approach to model-based control. Some of their representative tools include, but are not limited to, conformal prediction \cite{2022_Vovk_CP-book,2008_Shafer_CP-tutorial,2021_Angelopoulos_CP-Intro, 2024_Lindemann_CP-control-survey}, 
PAC-Bayes theory \cite{1998_PAC-Bayes}, 
scenario optimization \cite{2006_Scenario} (see also the references therein).
A thorough comparison among these methods in the context of data-driven controls can be found in \cite{2024_Lindemann_CP-control-survey}.
This paper focuses on conformal prediction (CP), a statistical tool based on rank statistics for quantifying uncertainties in data-driven prediction models \cite{2022_Vovk_CP-book,2008_Shafer_CP-tutorial,2021_Angelopoulos_CP-Intro}.
CP is well-suited to our problem setting because 1) it does not make any assumptions about the underlying distributions and structures of the uncertainties, 2) it is compatible with any prediction model, including neural networks \cite{2021_Angelopoulos_CP-Intro}, and 3) it requires only \emph{exchangeability} \cite{2022_Vovk_CP-book,2008_Shafer_CP-tutorial,2021_Angelopoulos_CP-Intro}, that is, the unseen real-world uncertainties are exchangeable with the empirically observed uncertainties. 
This assumption can be relaxed in some variations of CP \cite{2023_Barber_CP-BE, 2021_Gibbs_ACI}. Consequently, CP-based methods have demonstrated successes in various domains of uncertainty-aware planning and control \cite{2023_Lindemann_CP-Planning, 2023_Sun_CP-Diffusion, 2023_Dixit_ACP-Planning, 2023_Chee_WPC-MPC, 2024_Zhou_ACP-CBF-MPC, 2023_Yang_CP-Sensor}. For instance, CP has been used to obtain statistical coverage guarantees for data-driven dynamics models \cite{2023_Chee_WPC-MPC}, data-driven state estimation \cite{2023_Yang_CP-Sensor}, and environment prediction models \cite{2023_Lindemann_CP-Planning, 2023_Dixit_ACP-Planning}, to name a few, thereby constructing uncertainty-aware planning and safe control policies. 
In our proposed framework, we use CP to quantify model uncertainties over a finite time horizon, enabling its application in constructing stability and safety guarantees for continuous-time  nonlinear systems. 
\newpage

\section{Preliminaries}
\subsection{Closed-Loop Guarantees for Nonlinear Systems} \label{sec:guarantees}
Consider a nonlinear dynamical system given by 
\begin{align} \label{eq:true_dyn}
    \dot{x}=f(x, u)
\end{align}
where $x \in \Xcal \subset{\mathbb{R}^n}$ is the state vector, $u \in \Ucal\subset{\mathbb{R}^m}$ is the control input vector, and $f:\Xcal \times \Ucal \rightarrow \mathbb{R}^n$ is an \emph{unknown} locally Lipschitz function.

\begin{definition} [CLF \cite{2014_Ames_ES-CLF}] 
\label{def:CLF}
A continuously differentiable ($C^1$) function $V:\Xcal\to\mathbb{R}_{\geq0}$ is a control Lyapunov function (CLF) of \eqref{eq:true_dyn} on $\Xcal$ if $\exists c_1,c_2,c_3\in\mathbb{R}_{>0}$ s.t. $\forall x \in \Xcal$, 
\begin{align}
\label{eq:clf_cond_1}
    c_1\|x\|^2 \leq V(x) \leq c_2\|x\|^2
\end{align}
and 
\begin{align}
\label{eq:clf_cond_2}
    \inf_{u \in \Ucal} \left[\frac{\partial V}{\partial x}f(x,u) \right] \leq -c_3 V(x).
\end{align}
\end{definition}

\begin{lemma}[Exponential Stability \cite{2014_Ames_ES-CLF}]
The closed-loop system of \eqref{eq:true_dyn} with any locally Lipschitz control policy $$u(x)\in \left\{u \in \Ucal ~\middle\vert~ \frac{\partial V}{\partial x}f(x,u) \leq -c_3 V(x) \right\}$$ is exponentially stable at $x=0$. 
In other words,  we have $V(x(t)) \leq V(x(0)) e^{-c_3 t}$, and then 
\begin{align}
    \|x(t)\| \leq \sqrt{\frac{c_2}{c_1}}\|x(0)\| e^{-\frac{c_3}{2} t}
\end{align}
for all $x(0) \in \mathcal{D} \subseteq \Xcal$, where $\mathcal{D}$ is the region of attraction.
\end{lemma}

\begin{definition} [CBF \cite{2017_Ames_CBF-QP}] 
\label{def:CBF}
Let a safe set $\mathcal{C}\subset\Xcal$ be the 0-superlevel set of a $C^1$ function $h: \Xcal\rightarrow \mathbb{R}$, i.e., $\mathcal{C}=\{ x\in\Xcal | h(x)\geq 0\}$.
If there exists an extended class-$\mathcal{K}_\infty$ function\footnote{An extended class-$\mathcal{K}_\infty$ function is a continuous, strictly increasing function $\alpha: \mathbb{R}\rightarrow\mathbb{R}$ which satisfies $\alpha(0)=0$.} 
$\alpha$ 
such that $\forall x\in \Xcal$,
\begin{align} \label{eq:cbf_cond_2} 
     \sup_{u \in \Ucal}\left[\frac{\partial h}{\partial x}f(x,u)\right] \geq -\alpha(h(x))
\end{align}
then $h$ is a control barrier function (CBF) of \eqref{eq:true_dyn} on $\mathcal{C}$.
In this paper, we consider $\alpha(h(x))=\gamma h(x)$, where $\gamma\in\mathbb{R}_{>0}$.
\end{definition}

\begin{lemma} [Forward Invariance \cite{2017_Ames_CBF-QP}]
Any locally Lipschitz control policy $$u(x)\in \left\{u \in \Ucal ~\middle\vert~ \frac{\partial h}{\partial x}f(x,u) \geq -\gamma h(x) \right\}$$ renders $\mathcal{C}$ forward invariant, i.e., $$x(0)\in\mathcal{C}\Rightarrow x(t)\in\mathcal{C},~\forall t\in [0,\tau_{f}(x(0))]$$
where $[0,\tau_{f}(x(0))]$ is the maximal time interval where $x(t)$ is defined.
\end{lemma}

\begin{remark} \label{remark:existence_T}
    Since \eqref{eq:true_dyn} is locally Lipschitz, for each $x(0)\in\Xcal$, there exists a maximal time interval $[0,\tau_{f}(x(0))]$ where a unique solution exists. In this paper, we assume that there exists a finite time horizon $T\in\mathbb{R}_{>0}$ such that $\tau_{f}(x(0))\geq T,~\forall x(0)\in\Xcal$. In other words, a unique solution to \eqref{eq:true_dyn} exist on $[0,T],~\forall x(0)\in\Xcal$.
\end{remark}

\subsection{Finite-horizon Exponential Stability and Safety}
\label{sec:fh_stability_safety}
Extending the notion of finite-time stability in \cite{1965_Weiss_FT-Stability}, we define finite-horizon exponential stability as follows.

\begin{definition}
[Finite-horizon Exponential Stability]
\label{def:fh-exp_stability}
System \eqref{eq:true_dyn} is finite-horizon exponentially stable at $x=0$ in a finite time horizon $T\in\mathbb{R}_{>0}$ if there exists $U,W \in \mathbb{R}_{>0}$ such that
\begin{align}
    \|x(t)\| \leq U\|x(0)\| e^{-W t},~\forall t\in[0,T]\text{ and }\forall x(0) \in \mathcal{D}.
\end{align} 
\end{definition}

\begin{lemma} \label{lemma:fh-exp_stability}
Suppose that a $C^1$ function $V:\Xcal\to\mathbb{R}_{\geq0}$ satisfies \eqref{eq:clf_cond_1} and $V(x(t)) \leq V(x(0)) e^{-c_3 t},~\forall t\in[0,T]$ and $\forall x(0) \in \mathcal{D}$. 
Then, system \eqref{eq:true_dyn} is finite-horizon exponentially stable at $x=0$ over a horizon $T\in\mathbb{R}_{>0}$. In particular, we have $$\|x(t)\| \leq \sqrt{\frac{c_2}{c_1}}\|x(0)\| e^{-\frac{c_3}{2}t},~\forall t\in[0,T]\text{ and }\forall x(0) \in \mathcal{D}.$$ 
\end{lemma}

We also define finite-horizon safety as follows.
\begin{definition} \label{def:fh-safety}
[Finite-horizon Safety]
Given a safe set $\mathcal{C}$, system \eqref{eq:true_dyn} is safe in a finite time horizon $T\in\mathbb{R}_{>0}$ if the trajectories of \eqref{eq:true_dyn} satisfies the following:
\begin{align}
    x(0)\in\mathcal{C} \Rightarrow x(t) \in\mathcal{C},~\forall t\in[0,T].
\end{align}
\end{definition}

\begin{lemma} \label{lemma:fh-safety}
    Suppose that a safe set $\mathcal{C}$ is defined as in Definition \ref{def:CBF}, and there exists $\gamma\in\mathbb{R}_{>0}$ such that the $C^1$ function $h$ satisfies $h(x(t)) \geq h(x(0)) e^{-\gamma t},~\forall t\in[0,T]$ and $\forall x(0)\in\mathcal{C}$.
    Then, system \eqref{eq:true_dyn} is safe in the time horizon~$T$.
\end{lemma}

\subsection{Conformal Prediction} \label{sec:cp_theory}
Conformal prediction (CP) is a statistical method for quantifying uncertainty in data-driven prediction models.
Suppose we are given a regression model $\hat{Y}_i = \hat{f}(X_i)$ trained on a fixed training dataset $\Strain\coloneqq \left\{(X_i, Y_i) \right\}_{i\in\Itrain}$, where $X_i$ and $Y_i$ are features and labels, and $\Itrain=\{1,2,\dots \Ntrain\}$ is a set of indices.
To carry out CP, we randomly sample a calibration dataset $\Dcal\coloneqq \left\{(X_i, Y_i) \right\}_{i\in\Ical}$, where $\Ical=\{1,2,\dots \Ncal\}$ is a set of indices.
We require that $\Strain$ and $\Dcal$ are two disjoint sets.
Then, we define the \emph{nonconformity scores} as 
\begin{align} \label{eq:nonconformity_score}
    S_i \coloneqq \left\|\hat{f}(X_i)-Y_i\right\|
\end{align}
for all $i\in \Ical$.
Intuitively, smaller scores imply a more accurate regression model, and vice versa. 
Note that $\{S_i\}_{i\in\Ical}$ are random variables, where the randomness comes from the sampling of $\Dcal$.
Given a \emph{failure probability} $\delta\in(0,1)$, we define the \emph{conformal quantile} as the $(1-\delta)$-th quantile of the empirical distribution of $\{S_i\}_{i\in\Ical}\cup\infty$ \cite{2021_Angelopoulos_CP-Intro,2023_Barber_CP-BE}, denoted by
\begin{align} \label{eq:quantile}
    \quantile \coloneqq \mathrm{Quantile}\left(1-\delta; \sum_{i=1}^{\Ncal} \frac{1}{1+\Ncal} \delta_{S_i} + \frac{1}{1+\Ncal} \delta_{\infty}\right)
\end{align}
in which $\delta_a$ denotes the point mass probability at $a\in\mathbb{R}\cup\{-\infty, \infty\}$.
The quantity $\quantile$ can be equivalently interpreted as the $\lceil (1-\delta)(\Ncal+1)\rceil$-th smallest nonconformity score in $\{S_i\}_{i\in\Ical}\cup\{\infty\}$.
$\lceil\cdot\rceil$ denotes the ceiling function. 
Note that $\quantile$ is a random variable because $\{S_i\}_{i\in\Ical}$ are random variables.

\begin{lemma} [Marginal Coverage of CP \cite{2008_Shafer_CP-tutorial, 2021_Angelopoulos_CP-Intro, 2024_Lindemann_CP-control-survey}] 
\label{lemma:cp_marginal_coverage}
Suppose we have a regression model $\hat{Y}_i = \hat{f}(X_i)$ trained on a fixed $\Strain$.
For any new sample $(\XXnew, \YYnew)$, we define its nonconformity score $S_{i_{\mathrm{new}}}$ by \eqref{eq:nonconformity_score}.
If all the nonconformity scores in $\{S_i\}_{i\in\Ical\cup i_{\mathrm{new}}}$ are \emph{exchangeable}, then CP gives the following marginal coverage guarantee for the prediction error:
\begin{align}
    \Prob\left\{ \|\hat{f}(\XXnew)-\YYnew\| \leq \quantile ~\vert~  \Strain \right\} \geq 1-\delta,
\end{align}
where the probability is marginal over the randomness in the sampling of $\Dcal$ and $(\XXnew, \YYnew)$, conditioned on the fixed $\Strain$.
\end{lemma}

\begin{IEEEproof}
See Theorem 1 and Section 6.1 of \cite{2023_Barber_CP-BE}.
\end{IEEEproof}
The exchangeability condition in Lemma \ref{lemma:cp_marginal_coverage} means that the joint probability distributions of $\{S_i\}_{i\in\Ical\cup i_{\mathrm{new}}}$ and $\{s_{\sigma(i)}\}_{i\in\Ical\cup i_{\mathrm{new}}}$ are identical under any permutation $\sigma$.

\begin{remark} \label{remark:delta}
The statistical coverage in Lemma \ref{lemma:cp_marginal_coverage} is affected by the choice of $\delta$, but not $\Ntrain$ and $\Ncal$.
A smaller $\delta$ means a larger $\quantile$ (see \eqref{eq:quantile}), and vice versa.
However, note that $\Ncal < \lceil (1-\delta)(\Ncal+1)\rceil$ results in $\quantile=\infty$, which makes the coverage in Lemma \ref{lemma:cp_marginal_coverage} trivial. 
To obtain a meaningful coverage in Lemma \ref{lemma:cp_marginal_coverage}, \cite{2024_Lindemann_CP-control-survey} pointed out that we need $\Ncal \geq (1-\delta)/\delta$.
\end{remark}

\begin{lemma} [Conditional Coverage of CP \cite{2021_Angelopoulos_CP-Intro, 2024_Lindemann_CP-control-survey}]
\label{lemma:cp_conditional_coverage}
The coverage guarantee in Lemma \ref{lemma:cp_marginal_coverage} is said to be marginal because the probability in Lemma \ref{lemma:cp_marginal_coverage} is marginal over the sampling of $\Dcal$ and the new sample $(\XXnew, \YYnew)$.
If we fix the calibration set $\Dcal$ as well, i.e., conditioned on both $\Strain$ and $\Dcal$, then we have the conditional coverage guarantee given by 
\begin{align}
    \Prob\left\{\left\|\hat{f}(\XXnew)-\YYnew\right\| \leq \quantile~ \middle|~ \Strain\cup\Dcal \right\} \sim \mathrm{Beta}(k_\delta, \Ncal+1-k_\delta)
\end{align}
where $\mathrm{Beta}(\cdot,\cdot)$ denotes a beta distribution and $k_\delta=\lceil (1-\delta)(\Ncal+1)\rceil$.
Note that as $\Ncal\rightarrow\infty$, the mean of the beta distribution tends to $1-\delta$ and the variance tends to 0.
\end{lemma}

An insightful comparison between the marginal and conditional coverage guarantees of CP can be found in Section 2.1 of \cite{2024_Lindemann_CP-control-survey}.

To relax the exchangeability assumption, we could employ weighted conformal prediction \cite{2023_Barber_CP-BE}.

\begin{lemma} [Weighted CP\cite{2023_Barber_CP-BE}]
\label{lemma:weighted_cp}
First, assign custom weights $\{w_i\}_{i\in\Ical}$, where $w_i\in[0,1]$, to data in $\Dcal$.
Define the quantile as
\begin{align}
    \quantile_\mathrm{w} \coloneqq \mathrm{Quantile}\left(1-\delta; \sum_{i=1}^{\Ncal} \Tilde{w}_i\delta_{S_i} + \Tilde{w}_{\infty} \delta_{\infty}\right)
\end{align}
where $\tilde{w}_i = w_i/(\sum_{j=1}^{\Ncal}w_j + 1)$ and $\tilde{w}_{\infty} = 1/(\sum_{j=1}^{\Ncal}w_j + 1)$. 
Then, the statistical coverage becomes
\begin{align}
    \Prob\left\{\left\|\hat{f}(\XXnew)-\YYnew\right\| \leq \quantile_\mathrm{w}~\big\vert~\Strain \right\} \geq 1-\delta - \sum_{i=1}^{\Ncal} \tilde{w}_i \cdot d_{\mathrm{TV}} (Z, Z^i)
\end{align}
where $d_{\mathrm{TV}}(\cdot,\cdot)$ is the total variation distance between two distributions, $Z = \{S_i\}_{i\in\Ical\cup i_{\mathrm{new}}}$, and $Z^i$ is the score vector with $S_i$ swapped with $s_{i_\mathrm{new}}$ in $Z$. 
The weights $\{w_i\}_{i\in\Ical}$ are chosen to minimize the coverage gap $\sum_{i=1}^{\Ncal} \tilde{w}_i \cdot d_{\mathrm{TV}}(Z, Z^i)$, with higher weights assigned to data points similar in distribution to the new sample \cite{2023_Barber_CP-BE}. 
See \cite{2023_Barber_CP-BE, 2023_Chee_WPC-MPC} for details.
\end{lemma}

\section{Conformal Robustness} 
\label{sec:conformal_robustness}

In this paper, our objective is to construct finite-horizon exponential stability and safety guarantees for the unknown true system~\eqref{eq:true_dyn} using only a data-driven model. 
Specifically, suppose that a data-driven model $\hat{f}:\Xcal\times\Ucal\to\mathbb{R}^n$ is trained on a fixed training dataset $\Strain$. 
The model uncertainty is defined as
$$
    \Delta(x,u):=f(x,u)-\hat f(x,u).
$$
In Section \ref{sec:model_uq}, we will utilize conformal prediction in Section \ref{sec:cp_theory} to quantify this uncertainty over a finite horizon $T$.
Then, in Sections \ref{sec:CR-CLF} and \ref{sec:CR-CBF}, we will construct Lyapunov- and barrier-function-based controllers using $\hat{f}$ in conjunction with the CP-based uncertainty metric, such that the true system satisfies the desired finite-horizon guarantees with high probability.

\subsection{Uncertainty Quantification with Conformal Prediction} \label{sec:model_uq}
In this paper, we adopt the marginal type of coverage guarantee as in Lemma~\ref{lemma:cp_marginal_coverage}.
Suppose that we are given a data-driven model $\hat{f}$ trained on a fixed training dataset $\Strain$. 
To employ Lemma~\ref{lemma:cp_marginal_coverage}, we randomly sample a calibration dataset from \eqref{eq:true_dyn}, denoted by
\begin{align}
    \Dcald \coloneqq \left\{\phi_t(x_{0,i}), f(\phi_t(x_{0,i}), u(\cdot))\right\}_{i\in\Icald},
\end{align}
where $\Icald$ is a set of indices, and 
\begin{align}
   \phi_t(x_{0,i}) \coloneqq \phi(t; x_{0,i}, u(\cdot),T) 
\end{align}
denotes the trajectory that solves $\eqref{eq:true_dyn}$ in $t\in[0,T]$ with initial condition $x_{0,i}$ and under control policy $u(\cdot)$.
Here, $u(\cdot)\in\Ucal$ is any control input that renders $f(x,u(\cdot))$ locally Lipschitz, and $T\in\mathbb{R}_{>0}$ is a finite time horizon.
We define the non-conformity scores as 
\begin{align} \label{eq:nonconformity_score_traj}
    \Scored \coloneqq \sup_{t \in[0, T]} \left\|\Delta(\phi_t(x_{0,i}), u(\cdot)) \right\|
\end{align}
for all $i\in \Icald$.
Given a failure probability $\deltad\in(0,1)$, we define the conformal quantile for model uncertainty $\quantiled$ using \eqref{eq:quantile}.
For any newly sampled $(\phi_t(\Xonew), f(\phi_t(\Xonew),u(\cdot)))$, we also compute its nonconformity score $S_{i_{\mathrm{new}}}^\Delta$ using \eqref{eq:nonconformity_score_traj}. 
Note that $\{\Scored\}_{i\in\Icald\cup i_{\mathrm{new}}}$ and $\quantiled$ are random variables with randomness from the sampling of $\Dcald$ and the new sample.

\begin{remark} \label{remark:noise}
    We consider the case where $\Dcald$ sampled from \eqref{eq:true_dyn} is noise-free. 
    However, if measurement noise is present, the CP-based method in \cite{2023_Yang_CP-Sensor} can be incorporated into our framework to quantify the state estimation uncertainty.
\end{remark}

To utilize Lemma \ref{lemma:cp_marginal_coverage}, we make the following assumption on exchangeability. 
\begin{assumption} \label{assmp:exchangeability}
    For any newly sampled trajectory data $(\phi_t(\Xonew), f(\phi_t(\Xonew),u(\cdot)))$, 
    all the  nonconformity scores in $\{\Scored\}_{i\in\Icald\cup i_{\mathrm{new}}}$ are exchangeable.
    We do not assume $u(\cdot)$ to be the same across all samples. 
    We only assume $u(\cdot)$ renders $f(x,u(\cdot))$ locally Lipschitz continuous.
\end{assumption}

\begin{remark} \label{remark:still_exchangeable}
    To relax the exchangeability assumption, we could adopt weighted split CP (in Lemma \ref{lemma:weighted_cp}). This will introduce a coverage gap in the statistical coverage while the rest of our proposed theorems (to be presented in Sections \ref{sec:CR-CLF} and \ref{sec:CR-CBF}) remain unchanged. 
\end{remark}

The following corollary serves as our uncertainty quantification metric for the data-driven model.
\begin{corollary} \label{coro:cp_bound}
Suppose we are given a data-driven model $\hat{f}$ learned from $\Strain$. 
If Assumption \ref{assmp:exchangeability} holds for any newly sampled trajectory initialized at any $x(0)=\Xonew\in\Xcal$, then by Lemma \ref{lemma:cp_marginal_coverage}, 
\begin{align}
    \Prob\left\{\sup_{t \in[0, T]} \left\|\Delta(\phi_t(x(0)), u(\cdot))\right\| \leq \quantiled ~\middle|~ \Strain \right\} \geq 1-\deltad,
\end{align}
where the probability is marginal over the randomness in the sampling of $\Dcald$ and the newly sampled trajectory, conditioned on $\Strain$.
This also implies
\begin{align}
    \Prob\left\{\sup_{s \in[0, t]} \left\|\Delta(\phi_s(x(0)), u(\cdot))\right\| \leq \quantiled,~\forall t\in[0,T] ~\middle|~ \Strain \right\} \geq 1-\deltad.
\end{align}
\end{corollary}

In the rest of this paper, all the probability statements are conditioned on the fixed training dataset $\Strain$, and thus on the data-driven model $\hat{f}$.
For notation simplicity, we will write $\Prob\{\cdot\}\coloneqq\Prob\{\cdot\,|\,\Strain\}$.

\subsection{Conformally Robust Control Lyapunov Function} \label{sec:CR-CLF}
\begin{definition} [CR-CLF] \label{def:CR-CLF}
Suppose a data-driven model $\hat{f}$ is learned from the fixed training dataset $\Strain$. 
Let $\quantiled$ be a conformal quantile defined in \eqref{eq:quantile}, and denote $q^\Delta$ as a realization of $\quantiled$ once a calibration dataset $\Dcald$ is sampled.
A $C^1$ function $V:\Xcal\to\mathbb{R}_{\geq0}$ is a conformally robust CLF (CR-CLF) with respect to $\quantiled$ if it satisfies 
\eqref{eq:clf_cond_1}, and for every realization $q^\Delta$ of $\quantiled$,  
$\forall x\in\Xcal$, $\exists u\in\Ucal$ s.t.
\begin{align} \label{eq:crclf_cond}
    \frac{\partial V}{\partial x}\hat{f}(x,u) + c_3 V(x) + \left\|\frac{\partial V}{\partial x} \right\| q^\Delta \leq 0.
\end{align}
\end{definition}
\begin{theorem} [Statistical Finite-horizon Exponential Stability with CR-CLF]
\label{th:crclf_policy}
Suppose $\hat{f}$ is learned from the fixed $\Strain$. 
Suppose that $V$ is a CR-CLF with respect to $\quantiled$ as defined in Definition \ref{def:CR-CLF}.
Define a random admissible control set 
\begin{align}
    \mathbf{K}_{\mathrm{CL}}(x;\quantiled) \coloneqq \left\{u\in\Ucal ~\Big\vert~ \frac{\partial V}{\partial x}\hat{f}(x,u) + c_3 V(x)+\left\|\frac{\partial V}{\partial x}\right\| \quantiled \leq 0 \right\}.
\end{align}
For a realization $q^\Delta$ of $\quantiled$, we denote $\mathbf{K}_{\mathrm{CL}}(x;q^\Delta)$ as its path-wise realization. 
Suppose that Assumption \ref{assmp:exchangeability} holds for any newly sampled trajectory initialized at $x(0)\in \mathcal{D}$.
Recall that $\mathcal{D}$ is a region of attraction.
Then, any locally Lipschitz continuous control policy $u(x)\in \mathbf{K}_{\mathrm{CL}}(x;\quantiled)$ renders 
\begin{align}
\Prob\left\{ V(x(t)) \leq V(x(0)) e^{-c_3 t},~ \forall t\in[0,T] ~\middle|~ x(0)\in \mathcal{D} \right\} \geq 1-\deltad
\end{align}
and 
\begin{align} \label{eq:crclf_prob_guarantee}
    \Prob\left\{\|x(t)\| \leq \sqrt{\frac{c_2}{c_1}}\|x(0)\| e^{-\frac{c_3}{2}t}, ~\forall t\in[0,T] ~\middle|~ x(0)\in \mathcal{D} \right\} \geq 1-\deltad,
\end{align}
where the probability is marginal over the randomness in the sampling of $\Dcald$ and the newly sampled trajectory initialized at $x(0)\in \mathcal{D}$.
By Lemma~\ref{lemma:fh-exp_stability}, this implies that the closed-loop system is finite-horizon exponentially stable in the sense of Definition~\ref{def:fh-exp_stability} over the horizon $T$ with probability at least $1-\deltad$, marginal over $\Dcald$ and the newly sampled trajectory initialized at $x(0)\in\mathcal{D}$.
\end{theorem}

\begin{IEEEproof}
Define a coverage event
\begin{align}
    E^{\Delta} \coloneqq \left\{\sup_{t \in[0, T]} \left\|\Delta(\phi_t(x(0)), u(\cdot))\right\| \leq \quantiled \right\},
\end{align}
or equivalently 
\begin{align}
    E^{\Delta} = \left\{\sup_{s \in[0, t]} \left\|\Delta(\phi_s(x(0)), u(\cdot))\right\| \leq \quantiled,~\forall t\in[0,T] \right\},
\end{align}
where $x(0)\in \mathcal{D}$.

Fix any realization of $\Dcald$, and thus fix the corresponding realized $q^\Delta$ of $\quantiled$. 
For this realization, using~\eqref{eq:crclf_cond}, we have
\begin{align}
    \dot{V} &= \frac{\partial V}{\partial x} \left(\hat{f}(x, u) + \Delta(x,u)\right) \leq -c_3 V(x) - \left\|\frac{\partial V}{\partial x}\right\|q^\Delta +  \frac{\partial V}{\partial x} \Delta(x, u) \leq -c_3 V(x) + \left\|\frac{\partial V}{\partial x}\right\| (\| \Delta(x, u) \| - q^\Delta )\nonumber
\end{align}
where the second inequality is achieved by the Cauchy–Schwarz inequality. 
Multiplying $e^{c_3 t}$ on both sides of the inequality, we have
\begin{align}
    e^{c_3 t} \dot{V} + e^{c_3 t} c_3 V(x) \leq e^{c_3 t} \left\|\frac{\partial V}{\partial x}\right\| (\| \Delta(x, u) \| - q^\Delta).
\end{align}
Since $e^{c_3 t} \dot{V} + e^{c_3 t} c_3 V(x) = \frac{d}{dt}(e^{c_3 t} V(x))$, the inequality above can be written as
\begin{align}
    \frac{d}{dt}(e^{c_3 t} V(x)) \leq e^{c_3 t} \left\|\frac{\partial V}{\partial x}\right\| (\| \Delta(x, u) \| - q^\Delta ).
\end{align}
Integrating over time interval $[0,t]$, where $t\in[0,T]$, with $x(0) \in \mathcal{D}$ yields
\begin{align}
    e^{c_3 t} V(x(t)) - V(x(0)) \leq \scaleobj{0.8}{\int_{0}^{t}} e^{c_3\tau} \left\|\frac{\partial V}{\partial x}\right\|(\| \Delta(x, u) \| - q^\Delta) d\tau.
\end{align}
Rearranging the terms, we obtain 
\begin{align}
    &V(x(t)) 
    \leq V(x(0)) e^{-c_3 t}+\left(\sup_{s \in[0, t]} \| \Delta(\phi_s(x(0)), u(\cdot)) \| - q^\Delta \right)
    \scaleobj{0.8}{\int_{0}^{t}} e^{-c_3(t-\tau)} \left\|\frac{\partial V}{\partial x}\right\| d\tau
\end{align} 
for $t\in[0,T]$ and $x(0)\in \mathcal{D}$.
On the event $E^{\Delta}$, the second term on the right-hand side of the inequality above becomes non-positive, and thus we have 
$$V(x(t)) \leq V(x(0)) e^{-c_3 t}$$ 
for $t\in[0,T]$ and $x(0)\in \mathcal{D}$. 
Applying the bounds in \eqref{eq:clf_cond_1}, we further have
$$\|x(t)\| \leq \sqrt{\frac{c_2}{c_1}}\|x(0)\| e^{-\frac{c_3}{2}t}$$
for $t\in[0,T]$ and $x(0)\in \mathcal{D}$.

Suppose that Assumption \ref{assmp:exchangeability} holds for any newly sampled trajectory initialized at $x(0)\in \mathcal{D}$.
Then, applying the marginal guarantee in Corollary~\ref{coro:cp_bound}, we have that $\Prob(E^{\Delta} \mid x(0)\in\mathcal{D})\geq 1-\deltad$, of which the probability is marginal over the randomness of the sampling of $\Dcald$ and the newly sampled trajectory initialized at $x(0)\in \mathcal{D}$.
Thus, we can conclude that 
\begin{align}
    \Prob\left\{V(x(t)) \leq V(x(0)) e^{-c_3 t},~\forall t\in[0,T] ~\middle|~ x(0)\in \mathcal{D} \right\} \geq 1-\deltad
\end{align}
and
\begin{align}
    \Prob\left\{\|x(t)\| \leq \sqrt{\frac{c_2}{c_1}}\|x(0)\| e^{-\frac{c_3}{2}t}, ~\forall t\in[0,T] ~\middle|~ x(0)\in \mathcal{D} \right\} \geq 1-\deltad,
\end{align}
where the probability is marginal over $\Dcald$ and the newly sampled trajectory initialized at $x(0)\in \mathcal{D}$.

\end{IEEEproof}

Constructing a CR-CLF that exactly satisfies \eqref{eq:crclf_cond} may sometimes be challenging. Instead, our approximation of the CR-CLF might induce small violations of \eqref{eq:crclf_cond}. The following proposition analyzes the robustness of the CR-CLF against such violations. 
Note that we use $\Prob_{\quantiled}$ to denote the probability measure induced by the random variable $\quantiled$. 
We say that a property holds for $\Prob_{\quantiled}$-almost every realization $q^\Delta$ of $\quantiled$ if the set of realizations $q^\Delta$ for which the property fails has $\Prob_{\quantiled}$-measure zero.
Also, we say that an event occurs $\Prob_{\quantiled}$-almost surely if it occurs for all realizations of $\quantiled$ except possibly on a set of realizations with $\Prob_{\quantiled}$-measure zero.

\begin{proposition}[Approximated CR-CLF]
\label{prop:approx_crclf}
Let $\hat{f}$ be learned from $\Strain$.
Let $\quantiled$ denote the conformal quantile random variable computed by \eqref{eq:quantile}. 
After a calibration dataset $\Dcald$ is sampled, we denote $q^\Delta$ as the realized value of $\quantiled$.
A $C^1$ function $\hat V:\Xcal\to\mathbb{R}_{\geq0}$ is an approximated CR-CLF with respect to $\quantiled$ if it satisfies
\begin{align}
\label{eq:clf_cond_1_hat}
    c_1\|x\|^2 \leq \hat{V}(x) \leq c_2\|x\|^2
\end{align}
and for $\Prob_{\quantiled}$-almost every realization $q^\Delta$ of $\quantiled$, $\forall x\in\Xcal$, $\exists u\in\Ucal$ such that
\begin{align}
    \frac{\partial \hat V}{\partial x}\hat f(x,u)
    + c_3\hat V(x)
    +
    \left\|\frac{\partial \hat V}{\partial x}\right\| q^\Delta
    \le \rho(x;q^\Delta),
\end{align}
where $\rho(x;q^\Delta)\in\mathbb{R}_{\geq 0}$ denotes the violation associated with the realized quantile $q^\Delta$. 
Define a random admissible control set 
\begin{align}
    \mathbf{K}_{\mathrm{CL}}^{\rho}(x;\quantiled)
    \coloneqq \left\{
    u\in\Ucal
    ~\middle|~
    \frac{\partial \hat V}{\partial x}\hat f(x,u)
    + c_3\hat V(x)
    + \left\|\frac{\partial \hat V}{\partial x}\right\| \quantiled
    \leq \rho(x;\quantiled)
    \right\}.
\end{align}
For a realization $q^\Delta$ of $\quantiled$, we denote $\mathbf{K}_{\mathrm{CL}}^{\rho}(x;q^\Delta)$ as its path-wise realization. 

Next, we quantify the violations using a second layer of CP. 
For the realized quantile $q^\Delta$, we sample a second calibration dataset of trajectories under a locally Lipschitz continuous control policy $u(x)\in\mathbf{K}_{\mathrm{CL}}^{\rho}(x;q^\Delta)$:
\begin{align}
    \Dcalr(q^\Delta)
    \coloneqq
    \left\{
    \phi_t(x_{0,i})
    \right\}_{i\in\Icalr(q^\Delta)},
\end{align}
where $\Icalr(q^\Delta)$ denotes the set of indices. 
The corresponding violation scores are computed by
\begin{align}
    \Scorer(q^\Delta)
    \coloneqq
    \sup_{t\in[0,T]} \rho(\phi_t(x_{0,i}); q^\Delta),~\forall i\in\Icalr(q^\Delta).
\end{align}
Let $\quantiler(q^\Delta)$ denote the conformal quantile of $\{\Scorer(q^\Delta)\}_{i\in\Icalr(q^\Delta)}$ with a failure probability $\deltar\in(0,1)$.
Note that for a realization $q^\Delta$ of the first-layer
quantile $\quantiled$, $\quantiler(q^\Delta)$ is a random variable with randomness from the sampling of $\Dcalr(q^\Delta)$.
When the first-layer quantile $\quantiled$ is also random, we denote the second-layer quantile as $\quantiler(\quantiled)$.
After $\Dcald$ is sampled so that $\quantiled=q^\Delta$, and after $\Dcalr(q^\Delta)$ is sampled, we denote $q^\rho(q^\Delta)$ as the realized value of $\quantiler(q^\Delta)$.

For $\Prob_{\quantiled}$-almost every realization $q^\Delta$ of $\quantiled$, suppose that the violation score of any newly sampled trajectory initialized at $x(0)\in\mathcal D$ and under the policy $u(x)\in\mathbf{K}_{\mathrm{CL}}^{\rho}(x;q^\Delta)$ is exchangeable with $\{\Scorer(q^\Delta)\}_{i\in\Icalr(q^\Delta)}$. 
Also, suppose that the model-uncertainty score of the same newly sampled trajectory satisfies Assumption~\ref{assmp:exchangeability}. 
Then, for any newly sampled trajectory initialized at $x(0)\in\mathcal D$ and under control policy
$u(x)\in \mathbf{K}_{\mathrm{CL}}^{\rho}(x;\quantiled)$, we have
\begin{align}
    \Prob\left\{\hat V(x(t)) \leq \hat V(x(0))e^{-c_3t} + \frac{\quantiler(\quantiled)(1-e^{-c_3t})}{c_3},\ \forall t\in[0,T] ~\middle|~ x(0)\in \mathcal{D} \right\} \geq 1-\deltad-\deltar
\end{align}
and
\begin{align}
    \Prob\left\{ \|x(t)\| \leq \sqrt{ \frac{c_2}{c_1}\|x(0)\|^2e^{-c_3t} + \frac{\quantiler(\quantiled)(1-e^{-c_3t})}{c_1c_3}},\ \forall t\in[0,T] ~\middle|~ x(0)\in \mathcal{D} \right\} \geq 1-\deltad-\deltar.
\end{align}
The probabilities above are marginal over the first calibration dataset $\Dcald$ used to construct $\quantiled$, the second calibration dataset $\Dcalr(\quantiled)$ used to construct $\quantiler(\quantiled)$, and the same newly sampled trajectory on which both scores are evaluated.
They are not conditioned on a fixed realization of
$\Dcald$, $\quantiled$, $\Dcalr(\quantiled)$, or $\quantiler(\quantiled)$.

\end{proposition}

\begin{IEEEproof}
Define the model-uncertainty coverage event
\begin{align}
    E^{\Delta} \coloneqq \left\{ \sup_{s\in[0,t]} \left\| \Delta(\phi_s(x(0)),u(\cdot)) \right\| \leq \quantiled,\ \forall t\in[0,T] \right\},
\end{align}
where $x(0)\in\mathcal{D}$. 
By Corollary~\ref{coro:cp_bound},
\begin{align}
    \Prob(E^{\Delta} \mid x(0)\in\mathcal{D})\geq 1-\deltad,
\end{align}
where the probability is marginal over the sampling of $\Dcald$ and the newly sampled trajectory initialized at $x(0)\in\mathcal{D}$.
Next, define the violation coverage event
\begin{align}
    E^{\rho} \coloneqq \left\{ \sup_{s\in[0,t]} \rho(\phi_s(x(0));\quantiled) \leq \quantiler(\quantiled),\ \forall t\in[0,T] \right\},
\end{align}
where $x(0)\in\mathcal{D}$.
For $\Prob_{\quantiled}$-almost every $q^\Delta$, under the exchangeability assumption, conformal prediction gives
\begin{align}
    \Prob\left\{ \sup_{s\in[0,t]} \rho(\phi_s(x(0));q^\Delta) \leq \quantiler(q^\Delta),\ \forall t\in[0,T] ~\middle|~ \quantiled=q^\Delta, \,x(0)\in\mathcal{D} \right\} \geq 1-\deltar.
\end{align}
Equivalently, $\Prob(E^{\rho}\mid \quantiled, \,x(0)\in\mathcal{D}) \geq 1-\deltar$ $\Prob_{\quantiled}$-almost surely.
By the law of total probability, we have
\begin{align}
    \Prob(E^{\rho} \mid x(0)\in\mathcal{D}) 
    = \int_{\mathbb{R}} \Prob(E^{\rho}\mid \quantiled=q^\Delta, \,x(0)\in\mathcal{D})\, d\Prob_{\quantiled}(q^\Delta)
    = \mathbb E \left[ \Prob(E^{\rho}\mid \quantiled, \,x(0)\in\mathcal{D}) \right] \geq 1-\deltar.
\end{align}
Therefore, by the union bound,
\begin{align} \label{eq:approx_crclf_union_bound}
    \Prob(E^{\Delta} \cap E^{\rho} \mid x(0)\in\mathcal{D} ) \geq 1-\deltad-\deltar.
\end{align}

Now, fix an outcome in the joint event $E^{\Delta}\cap E^{\rho}$ with $x(0)\in\mathcal{D}$. 
For this outcome, the random quantile $\quantiled$ has a realized value $q^\Delta$, and the second-layer random quantile $\quantiler(q^\Delta)$ has a realized value $q^\rho(q^\Delta)$. 
For any locally Lipschitz continuous control policy
$u(x)\in\mathbf{K}_{\mathrm{CL}}^{\rho}(x;q^\Delta)$, we have
\begin{align}
\dot{\hat V} = \frac{\partial \hat V}{\partial x}
\left(\hat f(x,u)+\Delta(x,u)\right) &\leq
- c_3\hat V(x)
- \left\|\frac{\partial \hat V}{\partial x}\right\|q^\Delta
+ \rho(x;q^\Delta)
+ \frac{\partial \hat V}{\partial x}\Delta(x,u) \nonumber\\
&\leq
-c_3\hat V(x)
+\rho(x;q^\Delta)
+ \left\|\frac{\partial \hat V}{\partial x}\right\|
\left(\|\Delta(x,u)\|-q^\Delta\right),
\end{align}
where the second inequality follows from the Cauchy-Schwarz inequality. 
Multiplying both sides by $e^{c_3t}$ gives
\begin{align}
\frac{d}{dt}\left(e^{c_3t} \,\hat{V}(x(t))\right)
\leq e^{c_3t}\, \rho(x(t);q^\Delta) + e^{c_3t}
\left\|\frac{\partial \hat V}{\partial x}\right\|
\left(\|\Delta(x,u)\|-q^\Delta\right).
\end{align}
Integrating over $[0,t]$, where $t\in[0,T]$, with $x(0) \in \mathcal{D}$ and rearranging the terms yields
\begin{align}
\hat V(x(t))
\leq&
\hat V(x(0))e^{-c_3t}
+ \sup_{s \in[0, t]} \rho(\phi_s(x(0));q^\Delta) \int_0^t e^{-c_3(t-\tau)}\,d\tau \\
&+ \left(
\sup_{s\in[0,t]}\|\Delta(\phi_s(x(0)),u(\cdot))\|-q^\Delta
\right)
\int_0^t
e^{-c_3(t-\tau)}
\left\|\frac{\partial \hat V}{\partial x}\right\|
d\tau .
\end{align}
On the joint event $E^{\Delta}\cap E^{\rho}$, the realized values satisfy
\begin{align}
\sup_{s\in[0,t]} \|\Delta(\phi_s(x(0)),u(\cdot))\| \leq q^\Delta
\end{align}
and
\begin{align}
\sup_{s\in[0,t]} \rho(\phi_s(x(0));q^\Delta) \leq q^\rho(q^\Delta),
\end{align}
for every $t\in[0,T]$, which implies
\begin{align}
\label{eq:approx_crclf_realized_bound}
\hat V(x(t)) \leq \hat V(x(0))e^{-c_3t} + q^\rho(q^\Delta) \int_0^t e^{-c_3(t-\tau)}\,d\tau  
= \hat V(x(0))e^{-c_3t} + \frac{q^\rho(q^\Delta) (1-e^{-c_3t})}{c_3},
\end{align}
for $t\in[0,T]$ and $x(0) \in \mathcal{D}$.
Define an event
\begin{align}
E_V \coloneqq \left\{\hat V(x(t)) \leq \hat V(x(0))e^{-c_3t} + \frac{\quantiler(\quantiled) (1-e^{-c_3t})}{c_3},~\forall t\in[0,T]\right\},
\end{align}
where $x(0) \in \mathcal{D}$.
Note that every outcome in $E^{\Delta}\cap E^{\rho}$, which results in \eqref{eq:approx_crclf_realized_bound}, is also an outcome in $E_V$.
Therefore, $(E^{\Delta}\cap E^{\rho}) \subseteq E_V$, and thus 
$$\Prob\left\{ E_V \mid x(0)\in\mathcal{D}\right\} \geq \Prob\left\{ E^{\Delta}\cap E^{\rho} \mid x(0)\in\mathcal{D}\right\}.$$
By \eqref{eq:approx_crclf_union_bound}, we have
\begin{align}
    \Prob\left\{\hat V(x(t))\leq\hat V(x(0))e^{-c_3t} + \frac{\quantiler(\quantiled)(1-e^{-c_3t})}{c_3},\ \forall t\in[0,T] ~\middle|~ x(0)\in \mathcal{D}\right\} \geq 1-\deltad-\deltar. 
\end{align}
Because $\hat{V}$ satisfies \eqref{eq:clf_cond_1}, we further have
\begin{align}
    \Prob\left\{ \|x(t)\| \leq \sqrt{\frac{c_2}{c_1}\|x(0)\|^2e^{-c_3t}+\frac{\quantiler(\quantiled)(1-e^{-c_3t})}{c_1c_3}},\ \forall t\in[0,T] ~\middle|~ x(0)\in \mathcal{D}\right\} \geq 1-\deltad-\deltar.
\end{align}
\end{IEEEproof}

\begin{remark}
    Proposition \ref{prop:approx_crclf} does not guarantee finite-horizon exponential stability with high probability due to the term induced by the violations. 
    If $\quantiler(\quantiled)$ is bounded, then Proposition \ref{prop:approx_crclf} guarantees finite-horizon stability.
\end{remark}

\subsection{Conformally Robust Control Barrier Function} \label{sec:CR-CBF}
\begin{definition} [CR-CBF] \label{def:CR-CBF}
Suppose a data-driven model $\hat{f}$ is learned from the fixed training dataset $\Strain$. 
Let $\quantiled$ be a conformal quantile defined in \eqref{eq:quantile}, and denote $q^\Delta$ as a realization of $\quantiled$ once a calibration dataset $\Dcald$ is sampled.
A $C^1$ function $h:\Xcal\to\mathbb{R}$ is a conformally robust control barrier function (CR-CBF) with respect to $\quantiled$ if the safe set $\mathcal{C}$ is its 0-superlevel set, and for every realization $q^\Delta$ of $\quantiled$, $\forall x\in\Xcal,~\exists u\in\Ucal$ s.t.
\begin{align} \label{eq:crcbf_cond}
    \frac{\partial h}{\partial x} \hat{f}(x, u) + \gamma h(x) - \left\|\frac{\partial h}{\partial x}\right\| q^\Delta \geq 0.
\end{align}
\end{definition}

\begin{theorem} [Statistical Finite-horizon Safety with CR-CBF]
\label{th:crcbf_policy}
Suppose $\hat{f}$ is learned from the fixed $\Strain$. 
Suppose that $h$ is a CR-CBF with respect to $\quantiled$ as defined in Definition \ref{def:CR-CBF}.
Define a random admissible control set 
\begin{align}
    \mathbf{K}_{\mathrm{CB}}(x;\quantiled) \coloneqq \left\{u \in \Ucal ~\middle\vert~ \frac{\partial h}{\partial x} \hat{f}(x, u) + \gamma h(x)- \left\|\frac{\partial h}{\partial x}\right\| \quantiled \geq 0 \right\}.
\end{align}
For a realization $q^\Delta$ of $\quantiled$, we denote $\mathbf{K}_{\mathrm{CB}}(x;q^\Delta)$ as its path-wise realization. 
Suppose that Assumption \ref{assmp:exchangeability} holds for any newly sampled trajectory initialized at $x(0)\in \mathcal{C}$.
Then, any locally Lipschitz continuous control policy $u(x)\in \mathbf{K}_{\mathrm{CB}}(x;\quantiled)$ renders 
\begin{align}
    \Prob\left\{ x(t) \in \mathcal{C},~\forall t\in[0,T] ~\middle|~ x(0)\in\mathcal{C} \right\} \geq 1-\deltad,
\end{align}
where the probability is marginal over the randomness in the sampling of $\Dcald$ and the newly sampled trajectory initialized at $x(0)\in\mathcal{C}$.
By Lemma~\ref{lemma:fh-safety}, this implies that the closed-loop system is finite-horizon safe in the sense of Definition~\ref{def:fh-safety} over the horizon $T$ with probability at least $1-\deltad$, marginal over $\Dcald$ and the newly sampled trajectory initialized at $x(0)\in\mathcal{C}$.
\end{theorem}

\begin{IEEEproof}
Define a coverage event
\begin{align}
    E^{\Delta} \coloneqq \left\{\sup_{t \in[0, T]} \left\|\Delta(\phi_t(x(0)), u(\cdot))\right\| \leq \quantiled \right\},
\end{align}
or equivalently 
\begin{align}
    E^{\Delta} = \left\{\sup_{s \in[0, t]} \left\|\Delta(\phi_s(x(0)), u(\cdot))\right\| \leq \quantiled,~\forall t\in[0,T] \right\},
\end{align}
where $x(0)\in \mathcal{C}$.

Fix any realization of $\Dcald$, and thus fix the corresponding realized $q^\Delta$ of $\quantiled$. 
For this realization, using~\eqref{eq:crcbf_cond}, we have
\begin{align}
    \dot{h} = \frac{\partial h}{\partial x} (\hat{f}(x, u) + \Delta(x,u))\geq -\gamma h(x) + \left\|\frac{\partial h}{\partial x}\right\|q^\Delta + \frac{\partial h}{\partial x} \Delta(x, u)\geq -\gamma h(x) + \left\|\frac{\partial h}{\partial x}\right\| (q^\Delta - \| \Delta(x, u) \|)
\end{align}
where the second inequality is achieved by the Cauchy–Schwarz inequality. 
Multiplying $e^{\gamma t}$ on both sides, we have
\begin{align}
    e^{\gamma t} \dot{h} + e^{\gamma t}\gamma h(x) \geq e^{\gamma t} \left\|\frac{\partial h}{\partial x}\right\| (q^\Delta - \| \Delta(x, u) \|).
\end{align}
Since $e^{\gamma t} \dot{h} + e^{\gamma t}\gamma h(x)=\frac{d}{dt}e^{\gamma t} h(x)$, the inequality above becomes
\begin{align}
    \frac{d}{dt}e^{\gamma t} h(x) \geq e^{\gamma t} \left\|\frac{\partial h}{\partial x}\right\| (q^\Delta - \| \Delta(x, u) \|).
\end{align}
Integrating over $[0,t]$, where $t\in[0,T]$, with any initial state $x(0)\in\mathcal{C}$ yields
\begin{align}
    e^{\gamma t} h(x(t)) - h(x(0)) \geq \scaleobj{0.8}{\int_{0}^{t}} e^{\gamma \tau} \left\|\frac{\partial h}{\partial x}\right\| (q^\Delta - \| \Delta(x, u) \|) d\tau.
\end{align}
Rearranging the terms, we have
\begin{align}
    &h(x(t)) \geq e^{-\gamma t} h(x(0))- \left(\sup_{s \in[0, t]} \| \Delta(\phi_s(x(0)), u(\cdot)) \| - q^\Delta \right) \scaleobj{0.8}{\int_{0}^{t}} e^{-\gamma(t-\tau)} \left\|\frac{\partial h}{\partial x}\right\| d\tau
\end{align}
for $t\in[0,T]$ and $x(0)\in \mathcal{C}$.
On the event $E^{\Delta}$, the second term on the right-hand side of the inequality above becomes non-negative, and thus we have 
\begin{align}
    h(x(t)) \geq h(x(0)) e^{-\gamma t}
\end{align}
for $t\in[0,T]$ and $x(0)\in \mathcal{C}$.

Suppose that Assumption \ref{assmp:exchangeability} holds for any newly sampled trajectory initialized at $x(0)\in \mathcal{C}$.
Then, applying the marginal guarantee in Corollary~\ref{coro:cp_bound}, we have that $\Prob(E^{\Delta} \mid x(0)\in\mathcal{C})\geq 1-\deltad$, of which the probability is marginal over the randomness of the sampling of $\Dcald$ and the newly sampled trajectory initialized at $x(0)\in \mathcal{C}$.
Thus, we can conclude that 
\begin{align}
    \Prob\left\{h(x(t)) \geq h(x(0)) e^{-\gamma t}, ~\forall t\in[0,T] ~\middle|~ x(0)\in\mathcal{C} \right\} \geq 1-\deltad,
\end{align}
where the probability is marginal over $\Dcald$ and the newly sampled trajectory.
Therefore, by Lemma \ref{lemma:fh-safety}, we obtain
\begin{align}
    \Prob\left\{x(t) \in \mathcal{C},~\forall t\in[0,T] ~\middle|~ x(0)\in\mathcal{C} \right\} \geq 1-\deltad.
\end{align}

\end{IEEEproof}

As in CR-CLFs, sometimes we can only get an approximated CR-CBF with some violation in~\eqref{eq:crcbf_cond}. The robustness of the CR-CBF against such violations is discussed in the following proposition.

\begin{proposition} [Approximated CR-CBF]
\label{prop:approx_crcbf}
Let $\hat{f}$ be learned from $\Strain$.
Let $\quantiled$ denote the conformal quantile random variable
computed by \eqref{eq:quantile}. After a calibration dataset
$\Dcald$ is sampled, we denote $q^\Delta$ as the realized value of $\quantiled$. 
A $C^1$ function $\hat{h}:\Xcal\to\mathbb{R}$ is an approximated CR-CBF with respect to $\quantiled$ if the safe set $\mathcal{C}$ is its 0-superlevel set, and for $\Prob_{\quantiled}$-almost every realization
$q^\Delta$ of $\quantiled$, $\forall x\in\Xcal,~\exists u\in\Ucal$ s.t. 
\begin{align}
    \frac{\partial \hat{h}}{\partial x} \hat{f}(x, u) + \gamma \hat{h}(x) - \left\|\frac{\partial \hat{h}}{\partial x} \right\| q^\Delta \geq -\rho(x;q^\Delta),
\end{align}
where $\rho(x;q^\Delta)\in\mathbb{R}_{\geq 0}$ denotes the violation associated with
the realized quantile $q^\Delta$.
Define a random admissible control set
\begin{align}
    \mathbf{K}_{\mathrm{CB}}^{\rho}(x;\quantiled)
    \coloneqq
    \left\{
    u\in\Ucal
    ~\middle|~
    \frac{\partial \hat h}{\partial x}\hat f(x,u)
    + \gamma \hat h(x)
    - \left\|\frac{\partial \hat h}{\partial x}\right\|\quantiled
    \geq
    -\rho(x;\quantiled)
    \right\}.
\end{align}
For a realization $q^\Delta$ of $\quantiled$, we denote
$\mathbf{K}_{\mathrm{CB}}^{\rho}(x;q^\Delta)$ as its pathwise realization.

Next, we quantify the violations using a second layer of CP.
For the realized quantile $q^\Delta$, we sample a second calibration dataset of trajectories under a locally Lipschitz continuous control policy
$u(x)\in\mathbf{K}_{\mathrm{CB}}^{\rho}(x;q^\Delta)$:
\begin{align}
    \Dcalr(q^\Delta) \coloneqq \left\{ \phi_t(x_{0,i}) \right\}_{i\in\Icalr(q^\Delta)},
\end{align}
where $\Icalr(q^\Delta)$ denotes the set of indices. 
The corresponding violation scores are computed by
\begin{align}
    \Scorer(q^\Delta) \coloneqq \sup_{t\in[0,T]} \rho(\phi_t(x_{0,i});q^\Delta),~\forall i\in\Icalr(q^\Delta).
\end{align}
Let $\quantiler(q^\Delta)$ denote the conformal quantile of
$\{\Scorer(q^\Delta)\}_{i\in\Icalr(q^\Delta)}$ with failure probability
$\deltar\in(0,1)$. Note that for a realization $q^\Delta$ of the first-layer
quantile $\quantiled$, $\quantiler(q^\Delta)$ is a random variable with
randomness from the sampling of $\Dcalr(q^\Delta)$. When the first-layer
quantile $\quantiled$ is also random, we denote the second-layer quantile as
$\quantiler(\quantiled)$.
After $\Dcald$ is sampled so that $\quantiled=q^\Delta$, and after $\Dcalr(q^\Delta)$ is sampled, we denote $q^\rho(q^\Delta)$ as the realized value of $\quantiler(q^\Delta)$.

For $\Prob_{\quantiled}$-almost every realization $q^\Delta$ of $\quantiled$, suppose that the violation score of any newly sampled trajectory initialized at $x(0)\in\mathcal{C}$ and under the policy $u(x)\in\mathbf{K}_{\mathrm{CB}}^{\rho}(x;q^\Delta)$ is exchangeable with $\{\Scorer(q^\Delta)\}_{i\in\Icalr(q^\Delta)}$. 
Also, suppose that the model-uncertainty score of the same newly sampled trajectory satisfies Assumption~\ref{assmp:exchangeability}.
Then, for any newly sampled trajectory initialized at $x(0)\in\mathcal{C}$ and under control policy $u(x)\in\mathbf{K}_{\mathrm{CB}}^{\rho}(x;\quantiled)$, we have
\begin{align}
    \Prob\left\{ \hat h(x(t)) \geq \hat h(x(0))e^{-\gamma t} - \frac{\quantiler(\quantiled)(1-e^{-\gamma t})}{\gamma},~\forall t\in[0,T] ~\middle|~ x(0)\in\mathcal{C} \right\} 
    \geq 1-\deltad-\deltar.
\end{align}
The probability above is marginal over the first calibration dataset $\Dcald$ used to construct $\quantiled$, the second calibration dataset $\Dcalr(\quantiled)$ used to construct $\quantiler(\quantiled)$, and the same newly sampled trajectory on which both scores are evaluated. 
It is not conditioned on a fixed realization of $\Dcald$, $\quantiled$, $\Dcalr(\quantiled)$, or $\quantiler(\quantiled)$.

\end{proposition}

\begin{IEEEproof}
    The proof of Proposition \ref{prop:approx_crcbf} follows the same logic as the proof of Proposition \ref{prop:approx_crclf}.
\end{IEEEproof}

\begin{remark}
Proposition~\ref{prop:approx_crcbf} does not guarantee finite-horizon safety with high probability, since the term induced by violation may make the lower bound on $\hat h(x(t))$ negative. 
If we can further assume
\begin{align}
    \hat h(x(0))e^{-\gamma t} - \frac{\quantiler(\quantiled)(1-e^{-\gamma t})}{\gamma} \geq 0
\end{align}
for $t\in[0,T]$ and $x(0)\in\mathcal{C}$, then we can obtain the finite-horizon safety guarantee as follows
\begin{align}
    \Prob\left\{x(t) \in \mathcal{C},~\forall t\in[0,T] ~\middle|~ x(0)\in\mathcal{C} \right\} \geq 1-\deltad-\deltar.
\end{align}
\end{remark}

\section{Case Studies}
We consider four benchmark nonlinear systems. 
Using only data and without knowing the true dynamics of these systems, we implement the CR-CLF and CR-CBF.
These case studies illustrate the path-wise deployment of the proposed controllers after the conformal quantile is realized.
In this section, we use the sparse identification of nonlinear dynamics (SINDy) algorithm \cite{2016_SINDY} for model learning. 
The data-driven model is parameterized by $$\dot{x} = \hat{f}(x,u) = \xi\Theta^\top(x, u)$$
where $\Theta^\top:\Xcal\times\Ucal\rightarrow \mathbb{R}^{M\times 1}$ is a regressor, consisting of $M$ locally Lipschitz continuous basis functions; $\xi\in\mathbb{R}^{n\times M}$ is a parameter matrix to be learned, of which each row is a 1-by-$M$ parameter vector.
Given a training dataset $\Strain$, $\xi$ is learned using sparse regression \cite{2016_SINDY}. 
If the learned model is control-affine, let it also be expressed as 
\begin{align}
    \dot{x}=\hat{f}(x,u)=\xi\Theta^\top(x, u)=\hat{a}(x)+\hat{b}(x)u.
\end{align}
Note that our proposed framework is prediction-method-agnostic, which means the data-driven model $\hat{f}$ is not restricted to the SINDy structure and can be obtained by any appropriate regression or system-identification methods, including neural networks.
%

\subsection{Example 1: Inverted pendulum with CR-CLF}
We utilize the CR-CLF to stabilize an inverted pendulum. 
Let the state $x=[\theta~\dot{\theta}]^\top \in\mathbb{R}^2$, where $\theta$ denotes the angular position of the pendulum, and control $u\in\mathbb{R}$ be the torque input.
The true dynamics of the inverted pendulum is $$\dot{x} = 
    \begin{bmatrix}
    \dot{\theta} \\
    - \frac{b\dot{\theta}}{I} + \frac{mgL\sin\theta}{2I} 
    \end{bmatrix}
    +
    \begin{bmatrix}
    0 \\
    -\frac{1}{I}
    \end{bmatrix} u$$
where $m=1$ kg and $L=1$ m are the mass and length of the pendulum, $b=0.01$ is the damping, and $I=mL^2/3$ is the moment of inertia.
%
For model learning, we choose 
\begin{align}
\Theta(x,u)
= 
\begin{bmatrix}
    \theta^i \dot{\theta}^{\,j}
\end{bmatrix}_{i,j\in\mathbb{Z}_{\geq 0},\; i+j\leq 5}
\otimes
\begin{bmatrix}
    1 & u
\end{bmatrix} 
\end{align}
where $\big[\theta^i \dot{\theta}^{\,j}\big]_{i,j\in\mathbb{Z}_{\geq 0},\; i+j\leq 5}$ consists of all monomials in $(\theta,\dot{\theta})$ up to total degree five.
Note that the resulting $\hat{f}$ is control-affine.
For one sampled calibration set, applying CP gives the realized quantile $q^\Delta=0.196$ for $T=5$ s and $\deltad=0.1$.

We construct the CLF candidate as $V= x^\top P x$, where
$P$ solves $A^\top P + PA+Q=0$. Here, $A$ is the Jacobian of the closed-loop system of $\hat{f}$ with feedback control $u=[8~5]x$ and $Q=c_3 I_{2\times 2}$ \cite{choi2020cbfclfhelper}, where $c_3=0.5$ is the exponential decay rate.
By construction, such $V$ satisfies \eqref{eq:clf_cond_1}. 
Since $\hat{f}(x,u)$ is control-affine, we can synthesize a control policy that satisfies \eqref{eq:crclf_cond} by solving the following quadratic program (QP) \cite{2017_Ames_CBF-QP}:
\begin{equation} \label{eq:crclf_qp}
    u_{\mathrm{CR-CLF}}^\star(x) = \arg \min_{u \in \mathbf{K}_{\mathrm{CL}}(x;q^\Delta)} \, \| u \|^2.
    \tag{CR-CLF QP}
\end{equation}
For comparison, we also implement a regular, uncertainty-agnostic CLF QP controller $$u_{\mathrm{CLF}}^\star(x) = \arg \min_{u \in \mathbf{K}_{\mathrm{L}}(x)} \|u\|^2$$
where $\mathbf{K}_{\mathrm{L}}(x)=\{u \in \Ucal \mid \tfrac{\partial V}{\partial x}\hat{f}(x,u) + c_3 V(x) \leq 0 \}$.

\begin{proposition} \label{prop:Lipschitz_crclf_qp}
If $\hat{a}(x)$, $\hat{b}(x)$, and $\frac{\partial V}{\partial x}(x)$ are locally Lipschitz continuous, and $\frac{\partial V}{\partial x}\hat{b}(x)\neq 0$, then $u_{\mathrm{CR-CLF}}^\star(x)$ is locally Lipschitz continuous (see \cite{2017_Ames_CBF-QP} for proof), thus Theorem \ref{th:crclf_policy} can be applied.   
\end{proposition}

\begin{figure}[h!]
    \centering
    \includegraphics[width=0.8\textwidth]{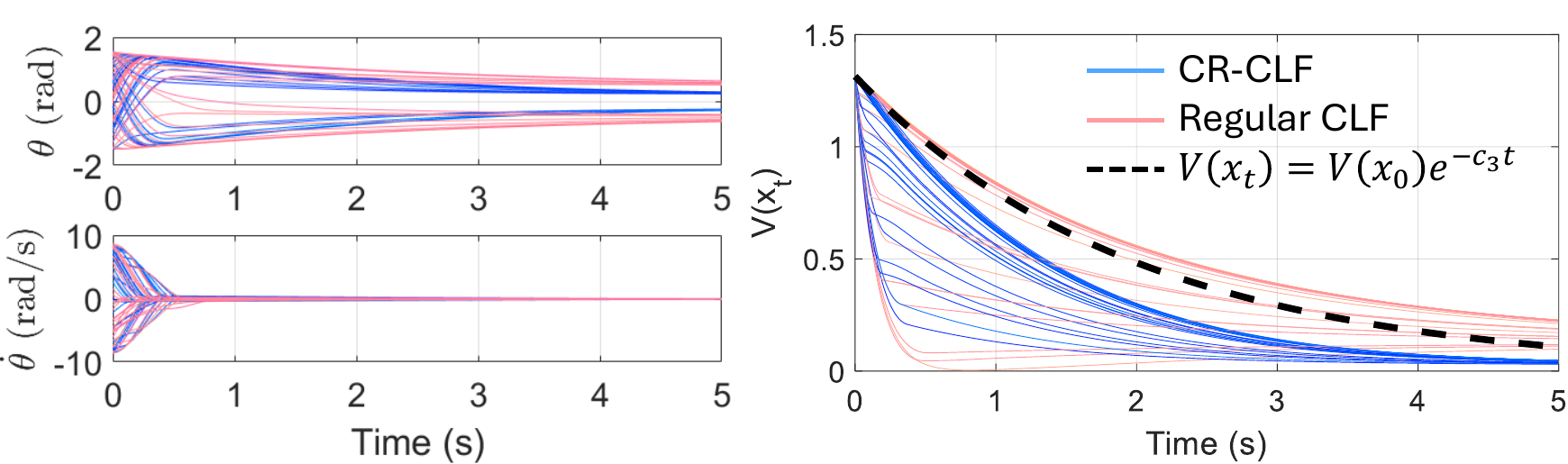}
    \caption{Inverted pendulum: comparison between the  CR-CLF and regular CLF.}
    \label{fig:ip}
\end{figure}

The code used for simulation is adapted from \cite{choi2020cbfclfhelper}.
For each controller, we simulate 30 trajectories starting from 30 random initial states in $\{x\in\Xcal\mid V(x) = 1.3\}$.
Fig. \ref{fig:ip} shows that $u_{\mathrm{CR-CLF}}^\star(x)$ given by \eqref{eq:crclf_qp} renders finite-horizon exponential stability of the true system for $T=5$ s, while $u_{\mathrm{CLF}}^\star(x)$ given by the regular CLF QP fails to achieve the same stability guarantee. 

\subsection{Example 2: Adaptive cruise control with CR-CBF}
We utilize the CR-CBF to guarantee finite-horizon safety for an adaptive cruise control (ACC) system \cite{2014_Ames_CBF-QP-ACC}, which consists of a leader and a follower vehicle. The follower is supposed to track a desired velocity while maintaining a safe distance from the leader. 
Let the state $x=[p~v~z]^\top\in\mathbb{R}^3$, where $p$ and $v$ denote the position and velocity of the follower, and $z$ denotes the distance between the leader and the follower. Let the control input $u\in\mathbb{R}$ be the force acting on the follower. The true system is given by $$\dot{x} = \begin{bmatrix}
    v \\
    -\frac{1}{m} (f_0 + f_1 v + f_2 v^2) \\
    v_0 - v
    \end{bmatrix} +
    \begin{bmatrix}
    0 \\
    \frac{1}{m}\\
    0
    \end{bmatrix} u$$
where $v_0=15$ m/s is the velocity of the leader; $m=2000$ kg is the mass of the follower; $f_0=0.5$, $f_1=5$, and $f_2=1$ are parameters \cite{2014_Ames_CBF-QP-ACC}.
For model learning, we choose $$\Theta(x,u)=\begin{bmatrix}1&p&v&z\end{bmatrix}\otimes\begin{bmatrix}1&u\end{bmatrix}.$$
Note that the resulting $\hat{f}$ is control-affine.
For one sampled calibration set, applying CP gives the realized quantile $q^\Delta=0.102$ for $T= 5$ s and $\deltad=0.05$.

We define the safe set $\mathcal{C} = \{(p,v,z) \mid z-T_h v \geq 0\}$, where $T_h=1$ s is the look-ahead time \cite{2014_Ames_CBF-QP-ACC}. 
A straightforward choice of the CBF candidate is $h(x)=z-T_h v$.
Since $\hat{f}(x,u)$ is control-affine, we can synthesize a control policy such that \eqref{eq:crcbf_cond} is satisfied by solving the following QP \cite{2017_Ames_CBF-QP}:
\begin{equation} \label{eq:crcbf_qp}
    u_{\mathrm{CR-CBF}}^\star(x) = \arg \min_{u \in \mathbf{K}_{\mathrm{CB}}(x;q^\Delta)} \, \| u -\overline{u}(x) \|^2,
    \tag{CR-CBF QP}
\end{equation}
in which $\overline{u}(x)$ is a reference control policy.
In this example, we set $\gamma=2$ in $\mathbf{K}_{\mathrm{CB}}(x;q^\Delta)$ and $\overline{u}=100(v_d-v)$, where $v_d=20$ m/s is the desired velocity.
For comparison, we also implement a regular, uncertainty-agnostic CBF QP controller by solving $$u_{\mathrm{CBF}}^\star(x) = \arg \min_{u \in \mathbf{K}_{\mathrm{B}}(x)} \| u -\overline{u}(x) \|^2$$
where $\mathbf{K}_{\mathrm{B}}(x)=\{u \in \Ucal \mid \tfrac{\partial h}{\partial x}\hat{f}(x,u) +\gamma h(x) \geq 0 \}$.

\begin{proposition} \label{prop:Lipschitz_crcbf_qp}
If $\hat{a}(x)$, $\hat{b}(x)$, and $\frac{\partial h}{\partial x}(x)$ are locally Lipschitz continuous, and $\frac{\partial h}{\partial x}\hat{b}(x)\neq 0$, then $u_{\mathrm{CR-CBF}}^\star(x)$ is locally Lipschitz continuous (see \cite{2017_Ames_CBF-QP} for proof), thus Theorem \ref{th:crcbf_policy} can be applied. 
\end{proposition}

\begin{figure}[h!]
    \centering
    \includegraphics[width=0.8\textwidth]{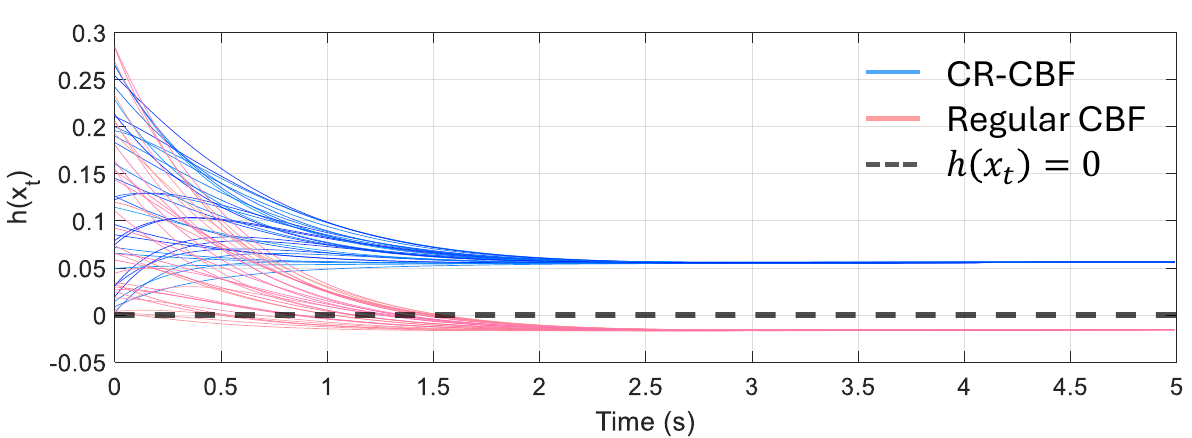}
    \caption{Adaptive cruise control: comparison between the CR-CBF and regular CBF.}
    \label{fig:acc}
\end{figure}

The code used for simulation is adapted from \cite{choi2020cbfclfhelper}.
We simulate 30 trajectories starting from 30 different initial states in the safe set $\mathcal{C}$.
Fig. \ref{fig:acc} shows that $u_{\mathrm{CR-CBF}}^\star(x)$ renders finite-horizon safety for $T=5$ s, while $u_{\mathrm{CBF}}^\star(x)$ renders unsafe behavior (i.e., $h(x(t))<0$).

\subsection{Example 3: Dubins car with neural CR-CBF}
We utilize a CR-CBF to maneuver a Dubins car towards a target while avoiding an obstacle. 
Let the state $x=[p_x~p_y~\theta]^\top\in\mathbb{R}^3$, where $p_x$ and $p_y$ are the $x$ and $y$ positions and $\theta$ is the heading. Let control input $u\in\mathbb{R}$ be the yaw rate.
The true model of the Dubins car is given by $$\dot{x} = \begin{bmatrix}v \cos\theta\\v \sin\theta\\0\end{bmatrix}+\begin{bmatrix}0\\0\\1\end{bmatrix} u$$
where $v=1$ m/s is a constant speed.
For model learning, we choose a regressor $$\Theta(x,u)=\begin{bmatrix}1&\theta~&\theta^2~&\theta^3\end{bmatrix}\otimes\begin{bmatrix}1&u\end{bmatrix}.$$
Note that the resulting $\hat{f}$ is control-affine.
For one sampled calibration set, applying CP gives the realized quantile $q^\Delta=0.554$ for $T=10$ s and $\deltad=0.1$.

Consider a circular obstacle with a radius of 2 m centered at the origin and a target at $(6,0)$ m. 
Starting from the left of the obstacle, the goal is to reach the target while avoiding the obstacle. 
We defined the safe set as $$\mathcal{C}=\left\{(p_x,p_y,\theta) ~\big\vert~ p_x^2+p_y^2 > 2^2,~|\theta| < \frac{\pi}{2} \right\}.$$ 
We adopt the framework in \cite{2023_Dawson_NCLBF} to learn the CR-CBF using neural networks. 
The neural CR-CBF, denoted as $h_{\theta_B}(x)$, is parameterized by $\theta_B$.
The learning is formulated as an empirical loss minimization problem:
$$\theta_B^\star = \arg\min_{\theta_B} \mathcal{L}_B.$$
The loss is defined by 
\begin{align}
    \mathcal{L}_B = \frac{\lambda_s}{N_s} \sum_{x_i\in\mathcal{C}} \mathrm{ReLU} \left(-h_{\theta_B}(x_i)\right) + \frac{\lambda_u}{N_u} \sum_{x_i\in\Xcal\setminus\mathcal{C}} \mathrm{ReLU} \left(h_{\theta_B}(x_i)\right) + \frac{\lambda_b}{N} \sum_{x_i\in\Xcal} \rho_i,
\end{align}
where $N=10^4$ denotes the total number of samples $x_i$ uniformly sampled from $\Xcal$; $N_{\mathrm{s}}$ (resp. $N_{\mathrm{u}}$) denotes the number of samples in the safe (resp. unsafe) set;
$\lambda_s=\lambda_u=100$ and $\lambda_b=1$ are the penalty weights;
$\rho_i$ is the slack variable obtained by solving 
\begin{align}
    \arg\min_{u\in\Ucal,\,\rho_i\in\mathbb R_{\geq 0}}
    \| u -\overline{u}(x_i) \|^2 + \lambda_\rho \rho_i
\end{align}
subject to
\begin{align}
    \frac{\partial h_{\theta_B}}{\partial x}\hat f(x_i,u)
    +\gamma h_{\theta_B}(x_i)
    -\left\|\frac{\partial h_{\theta_B}}{\partial x}\right\|q^\Delta
    \geq -\rho_i.
\end{align}
Here, $\lambda_\rho=10^4$, and $\overline{u}(x)=10\cdot \mathrm{wrapToPi}\bigl(\mathrm{atan2}(0-p_y, 6-p_x) - \theta\bigr)$ is a reference control policy.
This QP becomes identical to \eqref{eq:crcbf_qp} when $\rho_i \equiv 0$.
For comparison, we also learn a regular, uncertainty-agnostic neural CBF using the same method. For the regular neural CBF, $\rho_i$ is obtained by solving
\begin{align}
    \arg\min_{u\in\Ucal,\,\rho_i\in\mathbb R_{\geq 0}}
    \| u -\overline{u}(x_i) \|^2 + \lambda_\rho \rho_i
\end{align}
subject to
\begin{align}
    \frac{\partial h_{\theta_B}}{\partial x}\hat f(x_i,u)
    +\gamma h_{\theta_B}(x_i)
    \geq -\rho_i .
\end{align}

\begin{figure}[h!]
    \centering
    \includegraphics[width=0.8\textwidth]{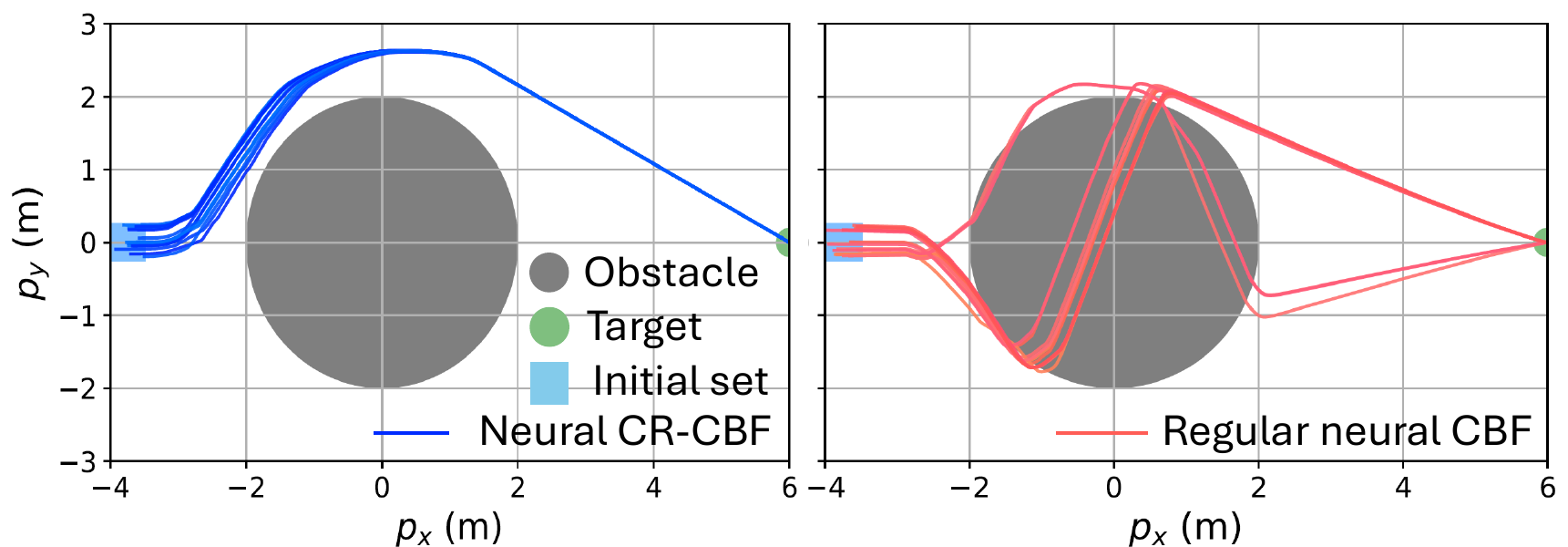}
    \caption{Dubins car collision avoidance: comparison between the neural CR-CBF and regular neural CBF.}
    \label{fig:dubnis}
\end{figure}

Fig. \ref{fig:dubnis} shows that the neural CR-CBF QP controller drives the Dubins car towards the target while avoiding the obstacle in the 10-second time horizon. However, the regular neural CBF QP controller results in collisions with the obstacle.

\subsection{Example 4: Cartpole with neural CR-CLF}
We utilize a CR-CLF to stabilize a cartpole at its upright position. 
Let the state $x=[z~\theta~\dot{z}~\dot{\theta}]^\top\in\mathbb{R}^4$, where $z$ is the position of the cart, and $\theta$ is the angular position of the pendulum. Let control input $u\in\mathbb{R}$ be the force acting on the cart.
The true system
is given by 
\begin{align}
\dot{x} =
\begin{bmatrix}
\dot{z} \\
\dot{\theta} \\
\frac{-m L \dot{\theta}^2 \sin\theta + 0.5 m g \sin(2\theta)}
{M + m \sin^2\theta} \\
\frac{-0.5 m L \dot{\theta}^2 \sin(2\theta) + (m + M) g \sin\theta}
{(M + m \sin^2\theta) L}
\end{bmatrix}
+
\begin{bmatrix}
0 \\
0 \\
\frac{1}{M + m \sin^2\theta} \\
\frac{\cos\theta}{(M + m \sin^2\theta) L}
\end{bmatrix} u
\end{align}
where $M=1$ kg and $m=0.3$ kg are the masses of the cart and the pendulum, and $L=1$ m is the length of the pendulum.
For model learning, we choose 
\begin{align}
\Theta(x,u)=&\begin{bmatrix}1&\dot{z}&\dot{\theta}&\dot{z}^2&\dot{z}\dot{\theta}&\dot{\theta}^2\end{bmatrix}\otimes\begin{bmatrix}1&\sin\theta&\cos\theta\end{bmatrix}\otimes\begin{bmatrix}1&\sin\theta&\cos\theta\end{bmatrix}\otimes\begin{bmatrix}1&\sin\theta&\cos\theta\end{bmatrix}\\
&\otimes\begin{bmatrix}1&\sin\theta&\cos\theta\end{bmatrix}\otimes\begin{bmatrix}1&u\end{bmatrix}.
\end{align}
Note that the resulting $\hat{f}$ is control-affine.
For one sampled calibration set, applying CP gives the realized quantile $q^\Delta=0.144$ for $T=5$ s and $\deltad=0.05$. 

We adapt the neural CLF in \cite{2023_Wei_NCLF-UU} to jointly learn the CR-CLF, $V_{\theta_V}(x)$, and its stabilizing controller, $u_{\theta_u}(x)$, using neural networks. 
$\theta_V$ and $\theta_u$ are the parameters of the neural networks to be learned. 
By construction in \cite{2023_Wei_NCLF-UU}, $V_{\theta_V}$ satisfies condition \eqref{eq:clf_cond_1} and $u_{\theta_u}(x)$ is locally Lipschitz continuous.
The learning of the CR-CLF is formulated as an empirical loss minimization problem: $$\theta_V^\star,\, \theta_u^\star = \arg\min_{\theta_V, \theta_u} \mathcal{L}_L.$$
We consider $\mathcal{L}_L = \mathcal{L}_d + \mathcal{L}_a$, in which 
$$\mathcal{L}_d = \frac{\lambda_d}{N} \sum_{i=1}^N \mathrm{ReLU}\left(\frac{\partial V_{\theta_V}}{\partial x}\hat{f}(x_i, u_{\theta_u}) + c_3 V_{\theta_V}(x_i) + \left\|\frac{\partial V_{\theta_V}}{\partial x}\right\| q^\Delta \right)$$
penalizes the violation of \eqref{eq:crclf_cond}, and 
$$\mathcal{L}_a = \frac{\lambda_a}{N} \sum_{i=1}^N \mathrm{ReLU} \left( -V_{\theta_V}(x_i) + \beta \|x_i\| \right)$$
encourages the growth of the region of attraction \cite{2019_Chang_NCLF}. 
$N$ denotes the number of samples. 
We set $\lambda_d=500$, $c_3 = 1$, $\lambda_a=300$, $\beta = 0.15$.
For comparison, we also learn a regular, uncertainty-agnostic neural CLF using the same method with $$\mathcal{L}_d = \frac{\lambda_d}{N} \sum_{i=1}^N \mathrm{ReLU}\left(\frac{\partial V_{\theta_V}}{\partial x} \hat{f}(x_i, u_{\theta_u}) + c_3 V_{\theta_V}(x_i) \right).$$

\begin{figure}[h!]
    \centering
    \includegraphics[width=0.8\textwidth]{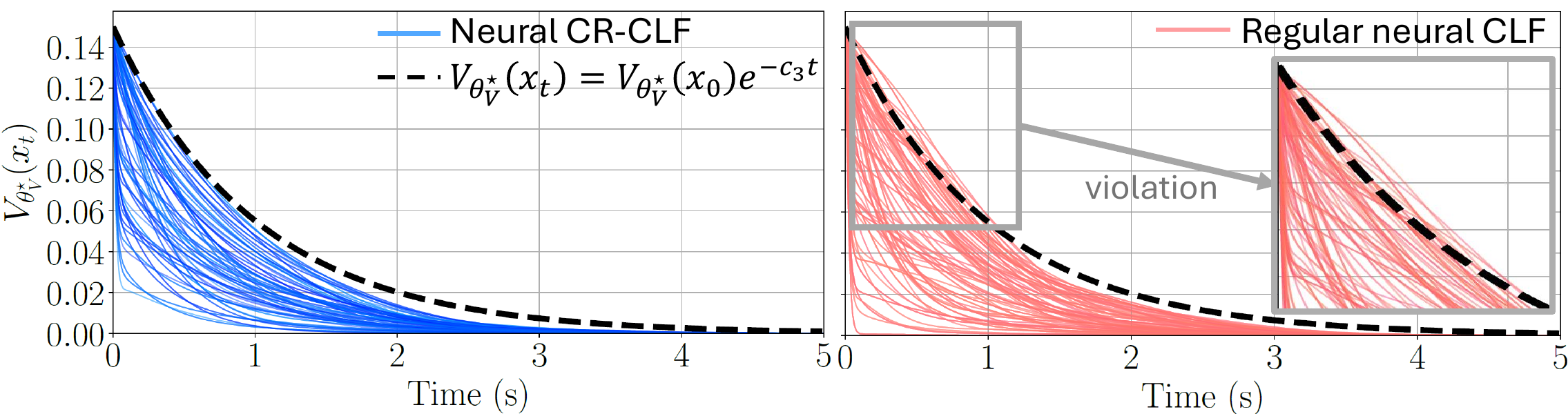}
    \caption{Cartpole: comparison between the neural CR-CLF and regular neural CLF.}
    \label{fig:cartpole}
\end{figure}

We simulate trajectories using both the neural CR-CLF and the regular neural CLF starting from 100 different initial states in $\{x\in\Xcal \mid V_{\theta_V}(x) = 0.15\}$. 
Fig. \ref{fig:cartpole} shows that the neural CR-CLF controller renders finite-horizon exponential stability for $T=5$ s, while the regular neural CLF controller results in violation of the upper bound on $V_{\theta_V^\star}(x(t))$. 

\section{Conclusion}
We introduced the concept of \emph{conformal robustness}. It provides a distribution-free, prediction-method-agnostic probabilistic bound on the model uncertainties for characterizing exponential stability and safety in fully data-driven, closed-loop, continuous-time nonlinear systems. We leveraged this concept to construct the CR-CLF and CR-CBF---explicit data-driven control designs for statistical finite-horizon guarantees of exponential stability and safety.
Simulations validated our framework, showing that the CR-CLF and CR-CBF synthesized by QP or neural networks outperformed the regular, uncertainty-agnostic CLF and CBF controllers in fully data-driven settings.

\newpage
\bibliographystyle{IEEEtran}
\bibliography{main}

\end{document}